\documentclass[epj]{svjour}
\usepackage{graphics}
\begin{document}
\title{Status of the $N^*$ Program at Jefferson Lab}
\author{Volker D. Burkert } 
%
%
\institute{Jefferson Lab, 12000 Jefferson Avenue, Newport News, VA23606}
\date{October 02, 2002}

%
\abstract{Recent results in electromagnetic excitation of nucleon  
resonance are presented, and confronted with theoretical predictions. 
Preliminary data in the search for missing states are discussed 
as well.}

\PACS{13.60.le, 13.88.+e} 
%
\maketitle

\section{Introduction}
\label{intro}

Resonance electroproduction has rich applications in nucleon structure 
studies at intermediate and large distances. Resonances play an 
important role in understanding the spin structure of the nucleon 
\cite{buli,ioffe}. More than 80\%  of the helicity-dependent
integrated total photoabsorption cross section difference (GDH integral) 
is a result of the excitation of the $\Delta(1232)$ \cite{buli,mami}. At 
$Q^2$= 1 GeV$^2$ about 40\% of the first moment 
$\Gamma_1^P(Q^2) = \int_0^1{g_1(x,Q^2)dx}$ for
the proton is due to contributions of the resonance region at 
$W < 2 $GeV \cite{devita,burkert_trieste,minehart}. 
Conclusions regarding the nucleon spin structure for $Q^2 < 2$ GeV$^2$ 
must therefore be regarded with some scepticism if contributions of 
baryon resonances are not taken into account. 

The nucleon's excitation spectrum has been explored mostly with pion beams. 
Many states, predicted in the standard quark model, have not been seen 
in these studies, possibly many of them decouple from 
the $N\pi$ channel \cite{koniuk}.
Electromagnetic interaction and measurement of multi-pion final states  
may then be the only way to study some of these states. 
While photoproduction is one way, electroproduction, though harder 
to measure, adds additional sensitivity due to the possibility of 
varying the photon virtuality.

\begin{figure}
\resizebox{0.48\textwidth}{!}{\includegraphics{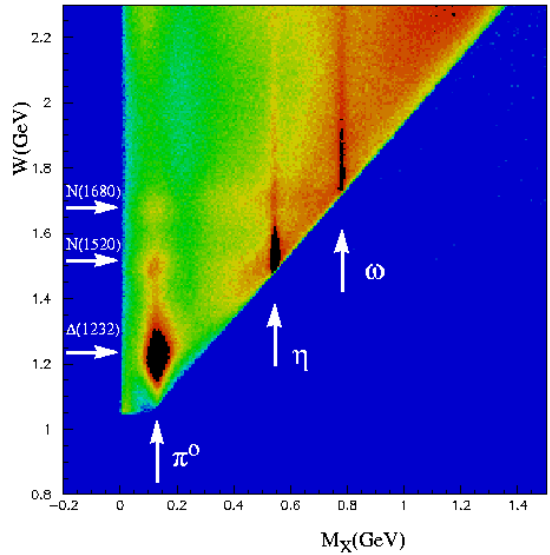}}
\caption{\small Hadronic invariant mass $W$ versus 
 missing mass $M_X$ for $\gamma^* p \rightarrow p X$, measured in CLAS. 
The vertical arrows indicate bands of $\pi^{\circ}$, $\eta$, and $\omega$ mesons.
The horizontal arrows mark the masses of several resonances.}
\label{fig:epx}
\end{figure}

Electroexcitation in the past was not considered a tool of 
baryon spectroscopy. CLAS is the first full acceptance instrument with 
sufficient resolution 
to measure exclusive electroproduction of mesons with the goal of studying 
the excitation of nucleon resonances in detail. The entire resonance
mass region, a large range in the photon virtuality $Q^2$ can be studied, 
and many meson final states are measured simultaneously. Figure \ref{fig:epx}
shows the coverage in the invariant hadronic mass W and the missing mass 
$M_X$ for the process $ep \rightarrow epX$ for a 4 GeV electron beam.

\begin{figure}
\resizebox{0.48\textwidth}{!}
{\includegraphics{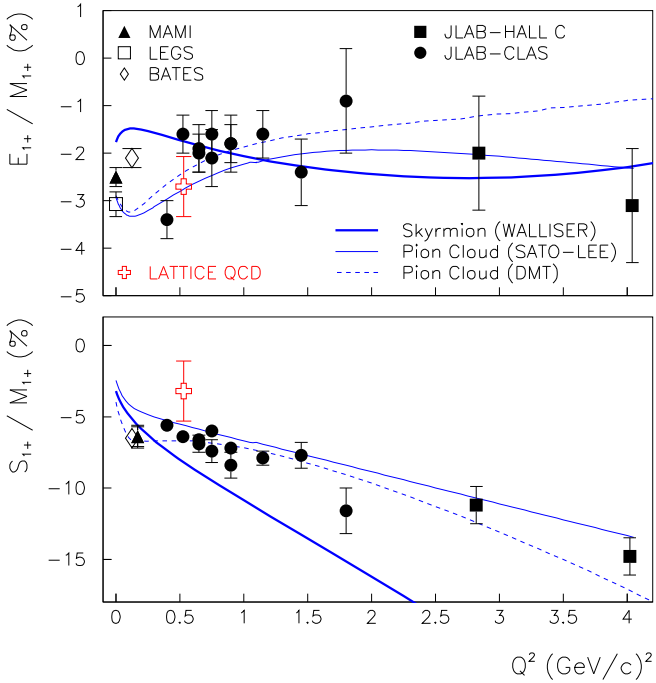}}
\caption{\small  $R_{EM}$ and $R_{SM}$ after 1990, including the recent CLAS 
results\cite{kjoo} and the data from Hall-C\cite{frolov}. It also shows the 
recent  Lattice QCD points.}
\label{remrsm}
\end{figure}
 
\section{Quadrupole deformation of the $\Delta(1232)$ and QCD}
\label{sec:ndelta}
An interesting aspects of nucleon structure at low energies 
is a possible quadrupole 
deformation of the nucleon or its lowest excited state. 
In the interpretation of ref. \cite{buchmann} 
this would be evident in non-zero values of the quadrupole transition 
amplitudes $E_{1+}$ and $S_{1+}$ from the nucleon to the $\Delta(1232)$.
In models with $SU(6)$ spherical symmetry, this transition is simply due to
a magnetic dipole $M_{1+}$ mediated by a spin flip 
from the $J= {1 \over 2}$ nucleon ground state to the $\Delta$ with 
$J = {3 \over 2}$, giving $E_{1+} = S_{1+} = 0$ . 
Non-zero values for $E_{1+}$ and $S_{1+}$ would indicate deformation. 
Dynamically such 
deformation may arise through interaction of the photon with the pion 
cloud\cite{sato,yang} or through the one-gluon 
exchange mechanism \cite{koniuk}. 
At asymptotic momentum transfer, a model-independent prediction of 
helicity conservation 
requires $R_{EM}\equiv E_{1+}/M_{1+} \rightarrow +1$. An interpretation 
of $R_{EM}$ in terms of a quadrupole deformation can therefore only be 
valid at low momentum transfer.

\begin{figure}[tb]
\resizebox{0.48\textwidth}{!}
{\includegraphics{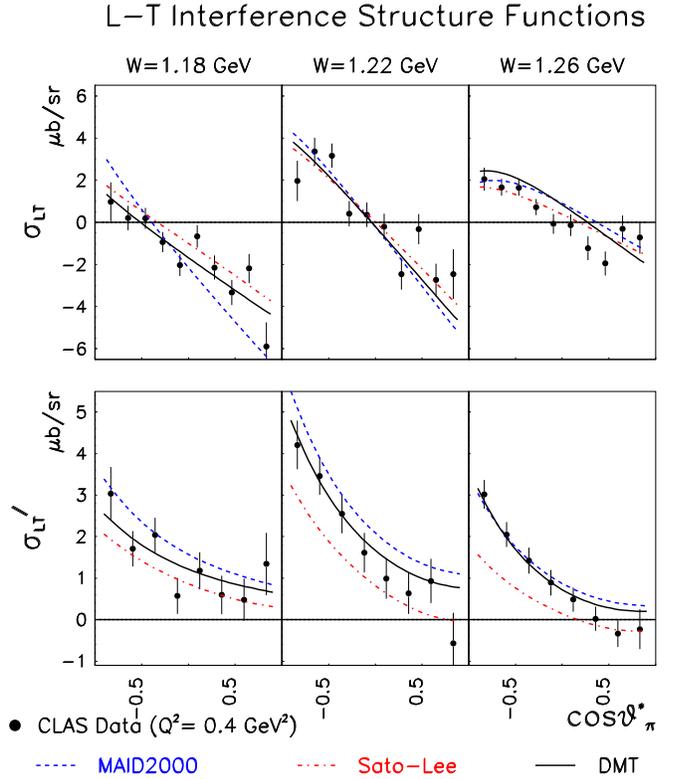}}
\caption{\small  Response functions $\sigma_{LT}$, and preliminary $\sigma_{LT^{\prime}}$ data for $\pi^o$ production from protons measured with 
CLAS\cite{kjoo,kjoo1} compared
to predictions of three dynamical models\cite{tiator,sato,yang}. 
The latter data show strong sensitivity to the non-resonant 
contributions in the various models.}
\label{fig:clasrltp}
\end{figure}

Results of the multipole analysis of the CLAS data\cite{kjoo} are shown
in Fig.\ref{remrsm}, where data from previous experiments 
published after 1990 are included as well \cite{beck,legs,frolov}. 
$R_{EM}$ remains negative and small throughout the $Q^2$ range. There are 
no indications that leading pQCD contributions are important as 
they would result in a rise of $R_{EM} \rightarrow +1$ \cite{carlson}. 
$R_{SM}$ behaves quite differently. While it also remains negative, 
its magnitude is strongly rising with $Q^2$. 
The comparison with microscopic models, from relativized quark 
models\cite{warns,aznaury}, the chiral quark soliton model\cite{silva}, and 
dynamical models\cite{sato,yang,dmt} show that simultaneous description of 
both $R_{EM}$ and $R_{SM}$ is achieved by dynamical models that include the 
nucleon pion cloud, explicitly. This supports the claim that most of the 
quadrupole strength is due to meson effects which are not included in 
other models.

Ultimately, we want to come to a QCD description of these important 
nucleon structure quantities.
At the time of this conference no lattice QCD calculations with sufficient 
accuracy were available to predict non-zero values for $R_{EM}$.
This situation has changed very recently with a calculation of the 
$R_{EM}$ and $R_{SM}$ ratio in quenched and unquenched QCD in the $Q^2$ 
range of the CLAS results\cite{alexandrou}. 
The full QCD results give $R_{EM}$ values more negative than in the 
quenched approximation showing the contribution of the pion cloud to be
negative, and causing an oblate deformation of the $\Delta(1232)$.
The calculation at $Q^2 = 0.52$ GeV$^2$ is in agreement 
with the CLAS data for $R_{EM}$ and $R_{SM}$.

While the new JLab data establish a new level of accuracy, improvements 
in statistics and the coverage of a larger $Q^2$ range are 
expected for the near future, and they must be complemented by a reduction 
of model dependencies in the analysis. 
This becomes increasingly important as the full QCD calculations get 
more precise, and calculations in a wide range of $Q^2$ may be 
forthcoming. It would be highly interesting to see if QCD calculations can 
describe the observed $Q^2$ evolution of $R_{SM}$.

\begin{figure*}
\vspace{65mm}
\centering{\includegraphics{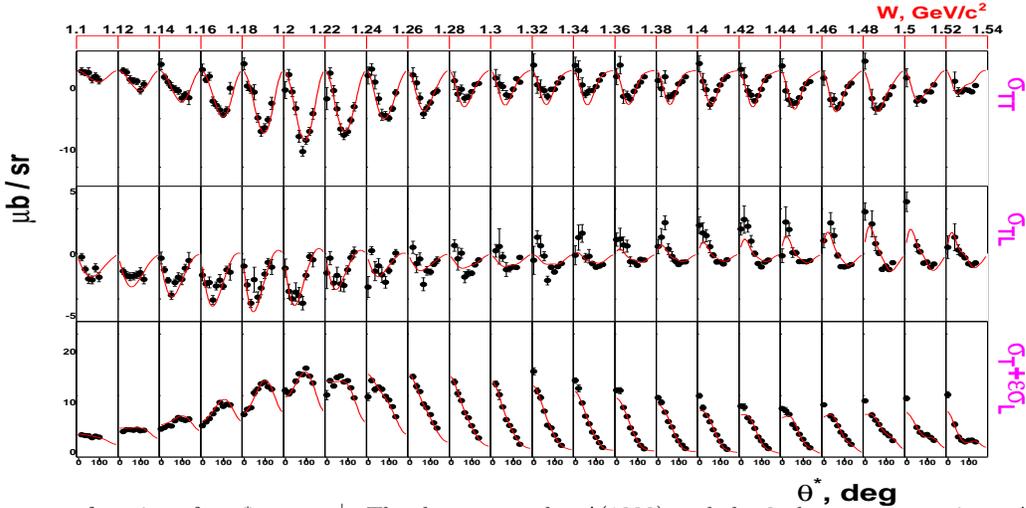}}
\caption{\small  Response functions for $\gamma^* p \rightarrow n\pi^+$. 
The data cover the $\Delta(1232)$ and the 2nd resonance regions. Angular 
distributions are show for each bin in W. The data provide the basis for
the analysis with a unitary isobar model\cite{janr}.}
\label{fig:npipl}
\end{figure*}

Model dependencies in the analysis are largely due to our poor knowledge 
of the non-resonant terms, which become increasingly important at higher 
$Q^2$. The 
$\sigma_{LT^{\prime}}$ response function, a longitudinal/transverse 
interference term is especially sensitive to 
non-resonant contributions if a strong resonance is present. 
$\sigma_{LT^{\prime}}$ can be measured using a polarized electron beam 
in out-of-plane kinematics for the pion. Preliminary 
data on $\sigma_{LT^{\prime}}$ from CLAS are shown in 
Fig. \ref{fig:clasrltp} in comparison with dynamical models, clearly showing
the model sensitivity to non-resonant contributions. 
All models predict nearly the same unpolarized 
cross sections at the $\Delta$ mass (upper panel for W = 1.22 GeV), 
however they differ in their handling of non-resonant contributions.

\section{$N^*$'s in the second resonance region}
\label{sec:2ndres}
Three states, the ``Roper'' $N^{\prime}_{1/2^+}(1440)$, 
and two strong negative parity states, $N^{*}_{3/2^-}(1520)$, 
and $N^{*}_{1/2^-}(1535)$ make up the 
second enhancement seen in inclusive electron scattering. All of these 
states are of special interest to obtain a better understanding of 
nucleon structure and strong QCD.

\subsection{The Roper resonance - still a mystery}

The Roper resonance has been a focus of attention for the last decade, largely due
to the inability of the standard constituent quark model to describe basic
features such as the mass, photocouplings, and their $Q^2$ evolution. 
This has led to alternate approaches where the state is assumed to have 
a strong gluonic component \cite{libuli}, a small quark core with a large 
meson cloud \cite{cano}, or a hadronic molecule of a nucleon and a 
hypothetical $\sigma$ meson $|N\sigma>$ \cite{krewald}.
Very recent lattice QCD calculations \cite{lattice-1} 
however indicate that the state may have a significant 3-quark component, 
and calculate the mass to be close to the experimental value.

Given these results the question what is the nature of
the existing Roper state becomes an urgent topic to address. 
Electroexcitation may help provide an answer as it probes the 
underlying structure. 

\begin{figure}
\vspace{65mm}
\centering{\includegraphics{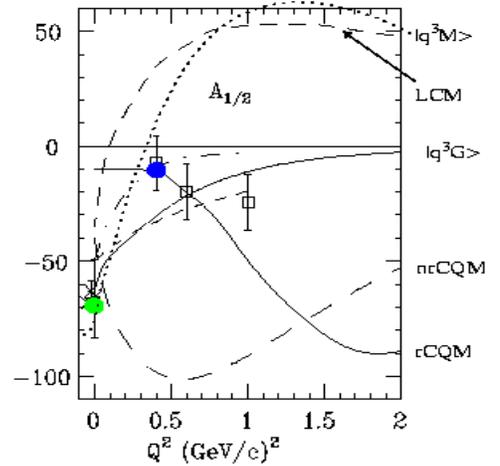}} 
\caption{\small Transverse helicity amplitude $A_{1/2}(Q^2)$ for the Roper 
resonance. The full (blue) circle shows preliminary results of an analysis 
of CLAS data at $Q^2 = 0.4$GeV$^2$. The curves represent model predictions.}
\label{fig:roper}
\end{figure}

The Roper, as an isospin 1/2 state, couples more strongly
to the n$\pi^+$ channel than to the p$\pi^o$ channel. Lack of data in
that channel and lack of polarization data has hampered progress in the past.
Fortunately, this sitation is changing significantly with the new data
from CLAS. For the first time complete angular distributions
have been measured for the n$\pi^+$ final state. Preliminary separated 
response functions obtained with CLAS\cite{hovanes} are shown in Fig.\ref{fig:npipl}.
These data, together with the $p\pi^o$  response functions and 
the spin polarized $\sigma_{LT^{\prime}}$ response function for both channels, 
have been fitted to a unitary isobar model\cite{janr}. 
The results are shown in Fig. \ref{fig:roper} together with the sparse 
data from previous analyses.  The CLAS results confirm the fast fall-off with 
$Q^2$ for the $A_{1/2}$ amplitude. Much improved data are needed for more 
definite tests of the models in a larger $Q^2$ range. 
An interesting question is whether the $A_{1/2}(Q^2)$ amplitude
changes sign, or remains negative. The range of model 
predictions for the $Q^2$ evolution illustrates dramatically the 
sensitivity of electroproduction to the internal structure of this 
state.

\subsection{The first negative parity state $N^*_{1/2^-}(1535)$}

Another state of special interest in the 2nd resonance region is the 
$N^*_{1/2^-}(1535)$. This state was found to have an unusually 
hard transition formfactor, i.e. the $Q^2$ evolution shows a slow
fall-off. This state is often studied in the $p\eta$ channel which
shows a strong s-wave resonance near the $\eta$-threshold with very little
non-resonant background. Older data show some discrepancies as to the 
total width and photocoupling amplitude. In particular, analyses of 
pion photoproduction data\cite{pdg} disagree with the analysis of
the $\eta$ photoproduction data by a wide margin.

\begin{figure}
\resizebox{0.48\textwidth}{!}{\includegraphics{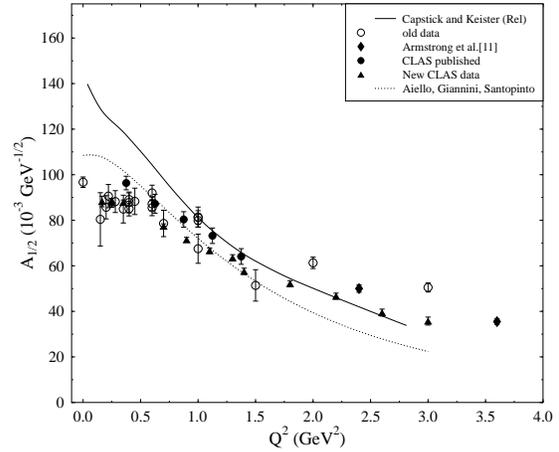}}
\caption{\small  Transverse helicity amplitude $A_{1/2}(Q^2)$ for the 
first negative parity state $N^*_{1/2^-}(1535)$.} 
\label{s11}
\end{figure}

Data from CLAS\cite{thompson}, together with 
data from an earlier JLab experiment\cite{armstrong} now give a 
consistent picture of 
the $Q^2$ evolution, confirming the hard formfactor behavior with 
much improved data quality, as shown in Fig. \ref{s11}.  Analysis of
the n$\pi^+$ and p$\pi^o$ data at $Q^2$= 0.4GeV$^2$ gives a value for 
$A_{1/2} \approx 105 \times 10^{-3}$ GeV$^{-1/2}$ consistent with the analysis of 
the p$\eta$ data\cite{aznaury-1}. 

The hard transition formfactor has been difficult to understand in models. 
Recent work  within a constituent quark model using a hypercentral potential 
\cite{santopinto} has made progress in reproducing the transition amplitude  
$A_{1/2}$ to the $N^*_{1/2^-}(1535)$. The hard formfactor is also
in contrast to models that interpret this state as a $|\bar K\Sigma>$
hadronic molecule \cite{weise}. Although no calculations 
exist from such models, the extreme ``hardness'' of the formfactor and the 
large cross section appear counter intuitive to an 
interpretation of this state as a bound hadronic system.
Lattice QCD calculations also show very clear 3-quark strength  
for the state \cite{lattice-1}.

\section{Higher mass states and missing resonances}
\label{sec:highermass}

Approximate $SU(6)\otimes O(3)$ symmetry of the symmetric constituent 
quark model leads to relationships between the various states. In the 
single-quark transition model (SQTM) only one quark participates in the 
interaction. The model predicts transition amplitudes for a large number of
states based on only a few measured amplitudes \cite{hey}. 
Comparison with photoproduction results show quite good agreement, while
there are insufficient electroproduction data for a meaningful 
comparison. The main reason for the lack of data on these states 
is that many of the higher mass states
decouple largely from the $N\pi$ channel, but couple dominantly to the 
$N\pi\pi$ channel. Study of $\gamma^*p \rightarrow p\pi^+\pi^-$ 
as well as the other charge channels are therefore important. 
Moreover, many of the so-called ``missing'' states are predicted to 
couple strongly to the $N\pi\pi$ channels \cite{capstick2}. 
Search for some of these states is of great importance for the understanding
of nucleon structure as alternative symmetry schemes do not predict 
nearly as many ``missing'' states\cite{kirchbach}.

\begin{figure}
\vspace{100mm}
\centering{\includegraphics{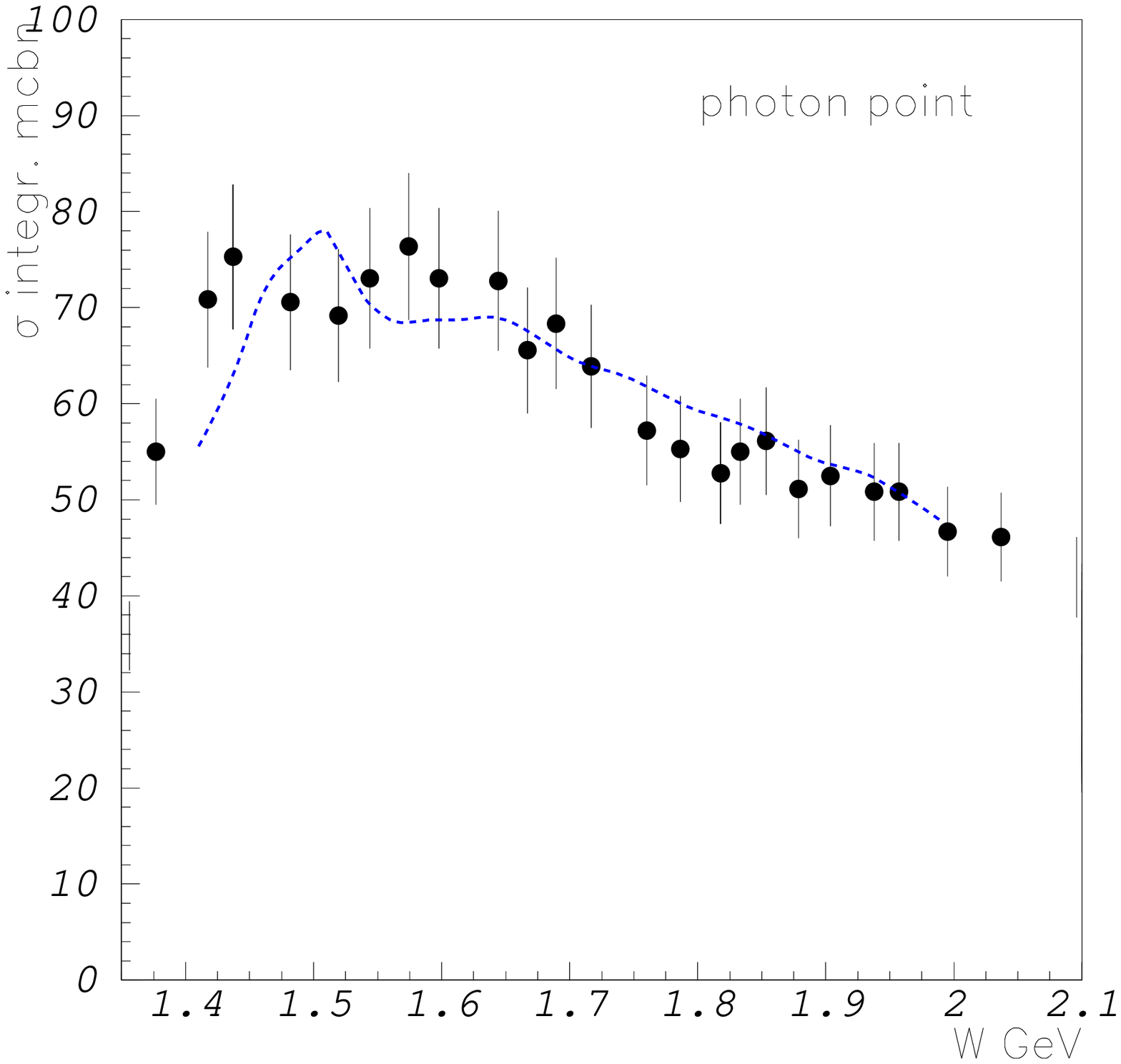}}
\centering{\includegraphics{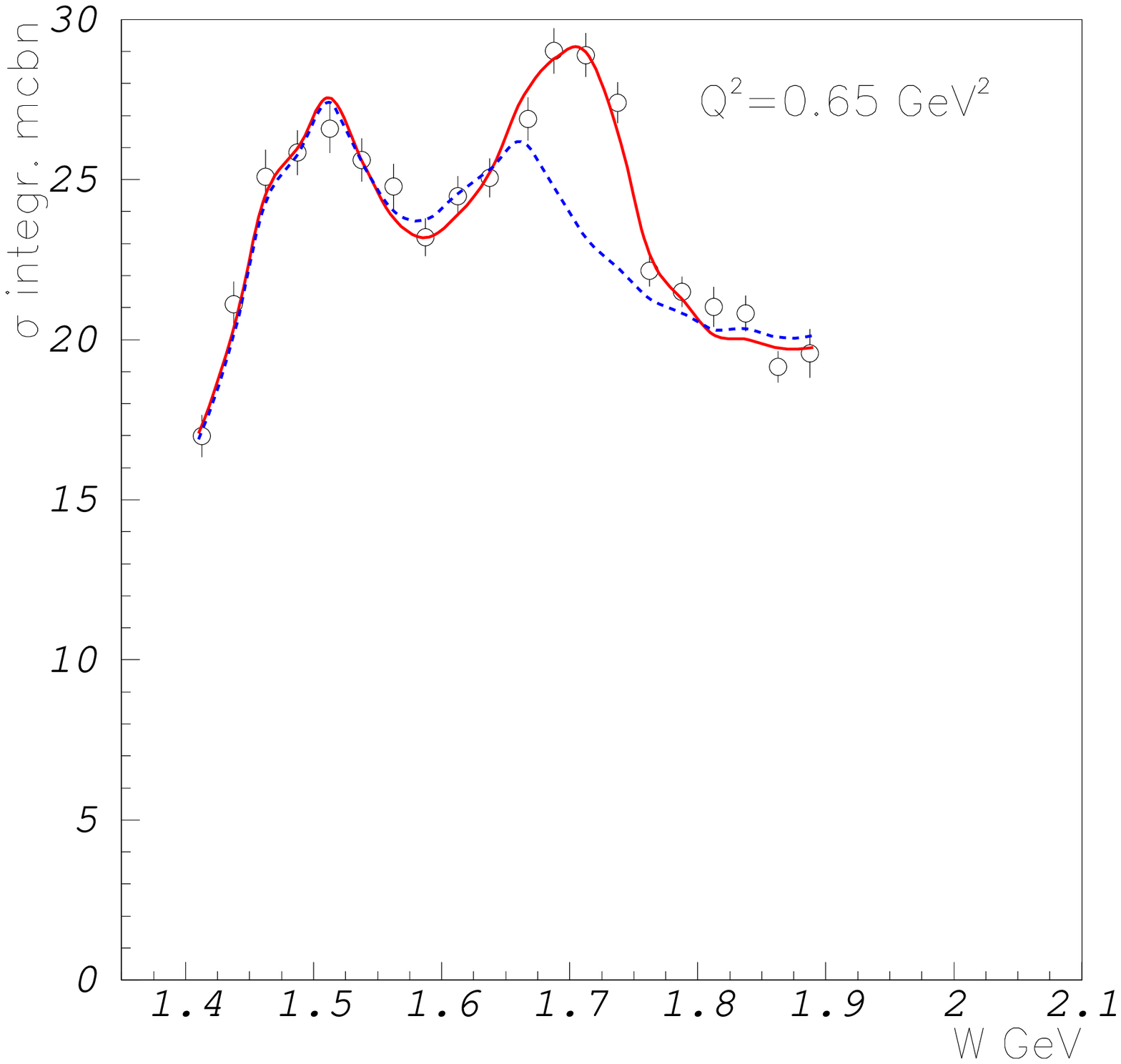}}
\centering{\includegraphics{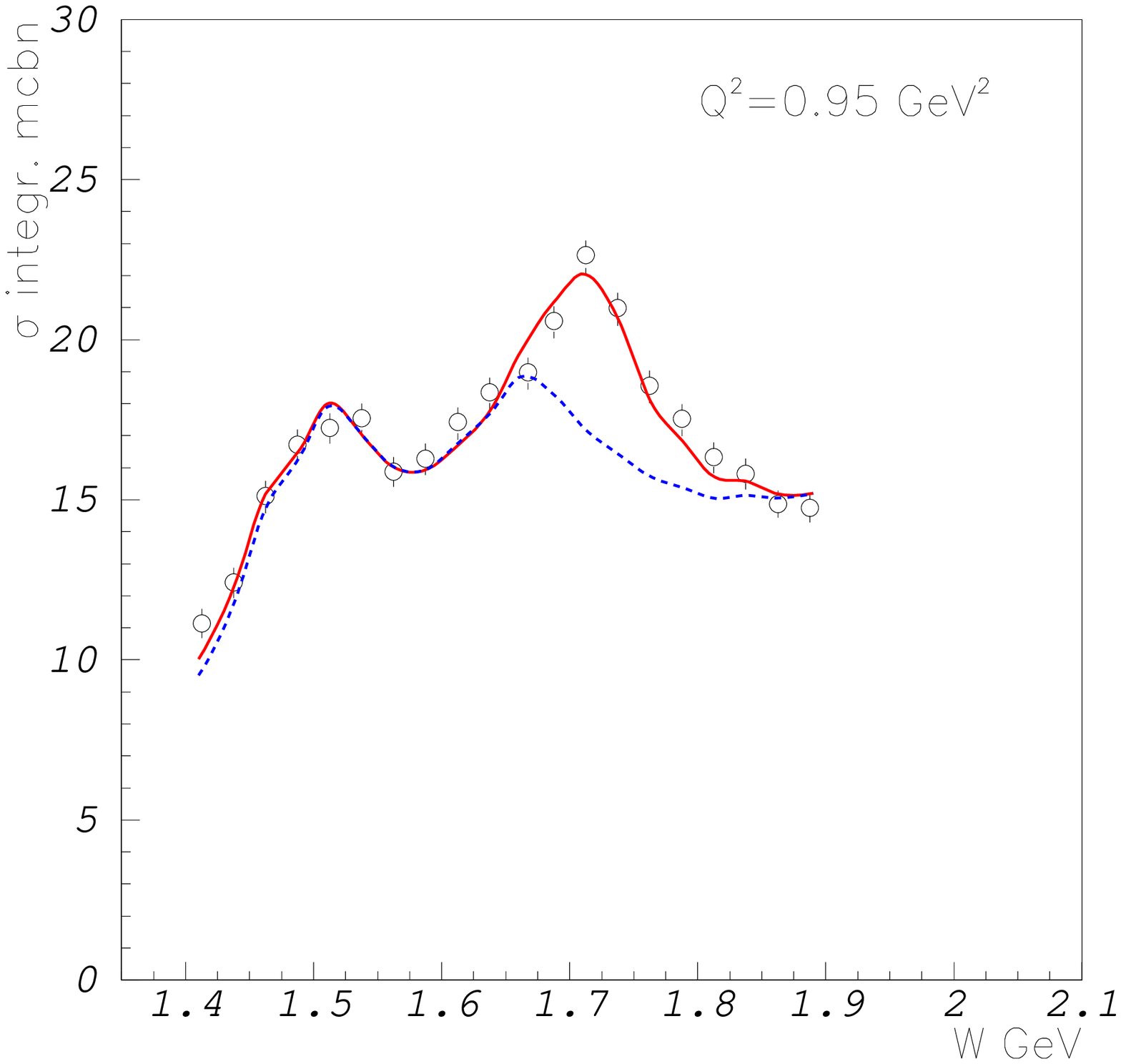}}
\centering{\includegraphics{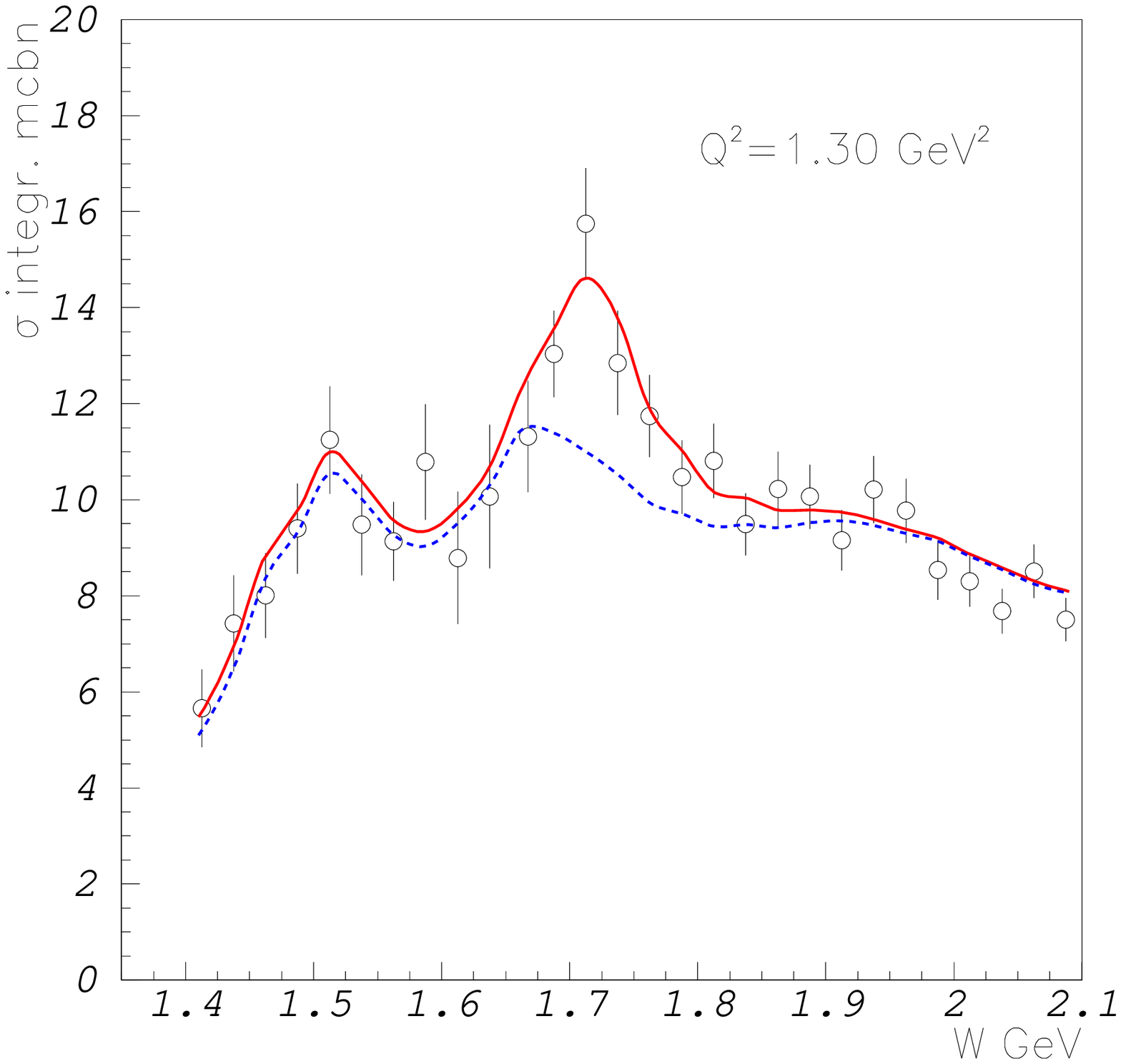}}
\caption{\small {Total photoabsorption cross section for 
$\gamma^* p \rightarrow p\pi^+\pi^-$. Photoproduction data from DESY - 
top left panel. The other panels show CLAS electroproduction data
at $Q^2=0.65, ~0.95, ~1.3$ GeV$^2$. The resonance structure near 1.7 GeV is emerging 
with increasing Q$^2$. The dashed lines represents our knowledge of $N^*$ electromagnetic
and hadronic properties with the couplings varied within empirical uncertainties. The 
solid line is a best fit to the data assuming the existence of a second  $N^*_{3/2^+}(1720)$ 
with different hadronic couplings.}}
\label{fig:ppippim}
\end{figure}

\begin{figure*}[tb]
\vspace{75mm}
\centering{\includegraphics{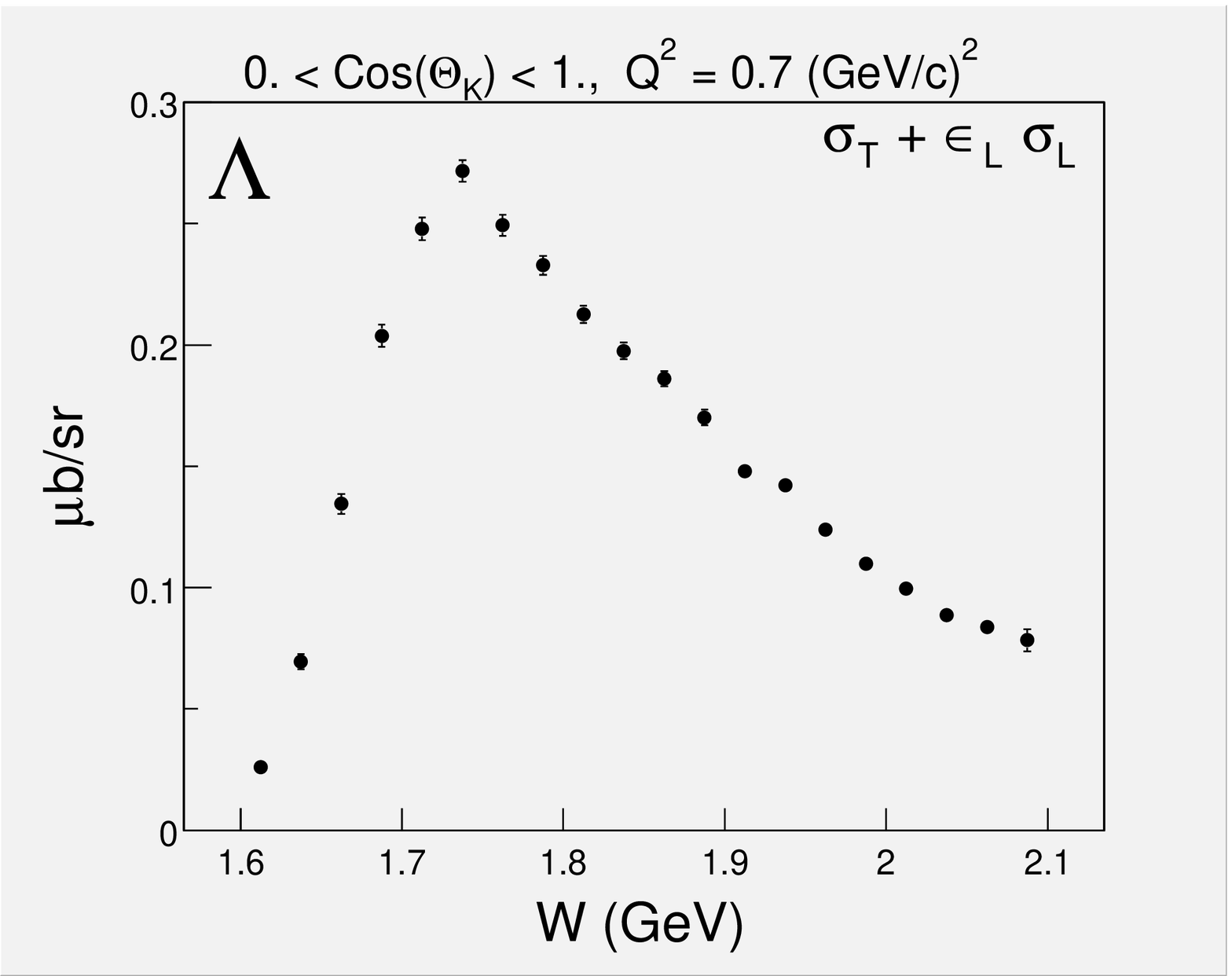}}
\centering{\includegraphics{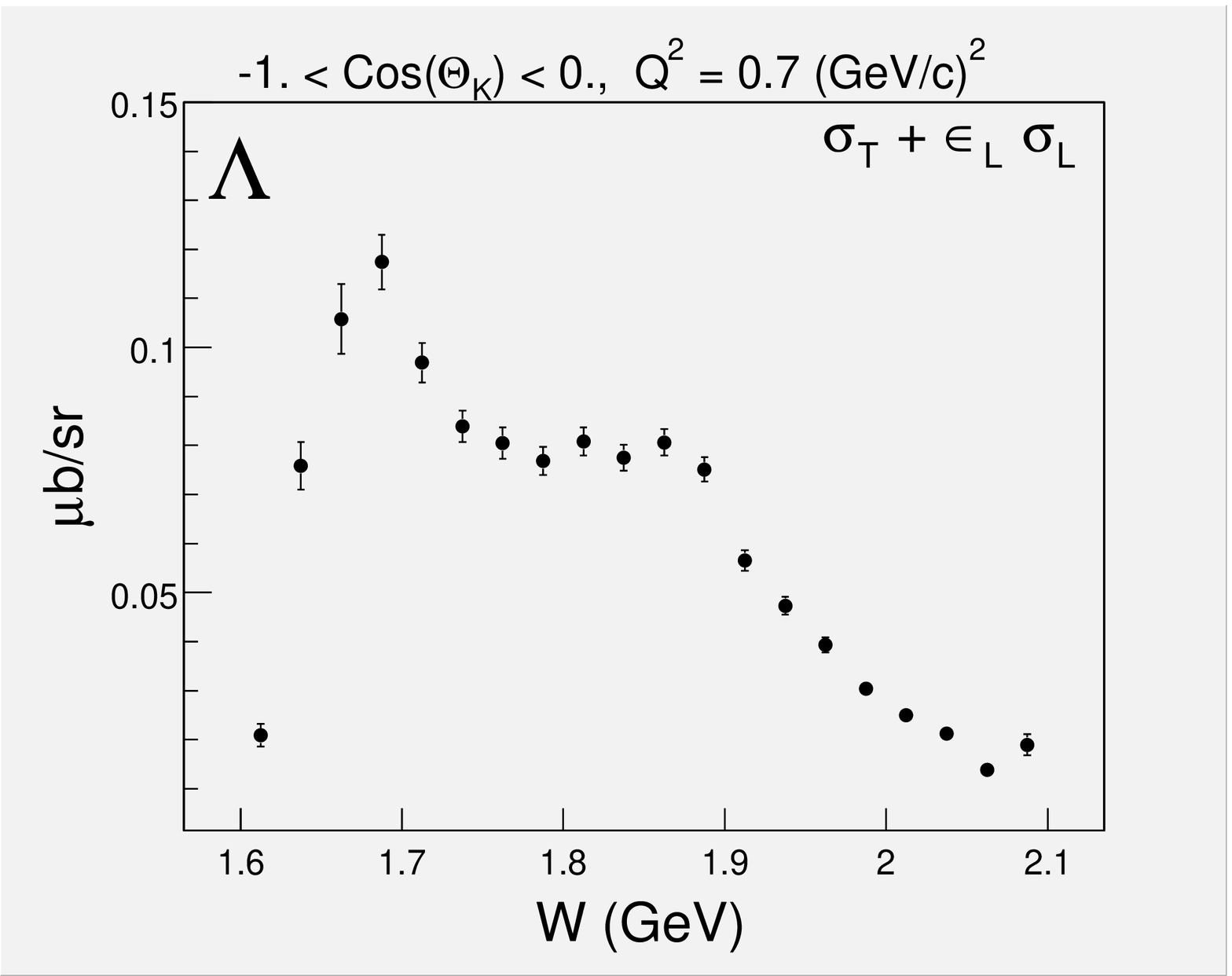}}
\caption{\small Total photoabsorption cross section measured with CLAS for 
$\gamma^* p \rightarrow K^+\Lambda$. The left panel is integrated over the full
forward hemisphere in the $K^+$ angular distribution in the $K^+\Lambda$ cms. 
The right panel is integrated over the backward hemisphere }.
\label{fig:klambda}
\end{figure*}

\subsection{ Resonances in the $p\pi^+\pi^-$ channel.}

New CLAS total cross section electroproduction data are shown in 
Fig. \ref{fig:ppippim} in comparison with photoproduction 
data from DESY \cite{desy}.
The most striking feature is the strong resonance peak near W=1.72 GeV
seen for the first time in electroproduction of the $p\pi^+\pi^-$ channel. 
This peak is absent in the photoproduction data. The CLAS 
data \cite{ripani} also contain the complete hadronic 
angular distributions and $p\pi^+$ and $\pi^+\pi^-$ 
mass distributions over the full W range. They have been analyzed and the
peak near 1.72 GeV was found to be best described by a $N^*_{3/2^+}(1720)$
state. While there exists a state with such quantum numbers in this mass range,
its hadronic properties were found previously to be very different from
the state observed in this experiment.
The difficulties in decribing these results seems to rest with the 
hadronic properties of the PDG state. Trying to keep the couplings within the 
limits of analyses of hadronic processes forces a strong reduction of the 
electrocouplings and the introducion of a second state with the same 
quantum numbers but strongly different hadronic couplings (solid line). 
 
Could this state be one of the ``missing'' states? 
Capstick and Roberts \cite{capstick2} predict a second  $N^*_{3/2^+}$
state at a mass 1.87GeV. There are also predictions of a hybrid baryon
state with these quantum numbers at about the same mass \cite{page}, although 
the rather hard form factor disfavors the hybrid baryon 
interpretation \cite{libuli}.
As mass predictions in these models are uncertain to at least
$\pm$100MeV, interpretation of this state as a ``missing'' state is a 
definite possibility. 
Independent of possible interpretations, the hadronic properties of the state 
seen in the CLAS data 
appear incompatible with the properties of the known state with same 
quantum numbers as listed in Review of Particle Properties \cite{pdg} and
the analyses of $N\pi\pi$ final states in $\pi N$ scattering. 

\subsection{Nucleon states in $K\Lambda$ production?}

Strangeness channels have recently been examined in photoproduction
as a possible source of information on new baryon states,
and candidate states have been discussed \cite{saphir,angelo}.
New CLAS electroproduction data  \cite{niculescu} in the $K\Lambda$ 
channel show clear evidence for resonance excitations at masses of 1.7 and 
1.85 GeV as show in Fig. \ref{fig:klambda}. The analysis of the 
$K\Lambda$ channel is somewhat complicated by the large t-channel 
exchange contribution producing a peak at forward angles.
To increase sensitivity to s-channel processes the data have been 
divided into a set for the forward hemisphere
and for the backward hemisphere. Clear structures in the invariant mass 
emerge for the backward hemisphere (right panel in Fig. \ref{fig:klambda}).
While the lower mass peak is probably due to known resonances, the peak 
near 1.85 GeV could be associated with the bump observed with 
the SAPHIR detector \cite{saphir}, although its mass seems to be lower. 
A more complete analysis of the angular distribution and the energy-dependence 
is needed for more definite conclusions.  

\subsection{Photoproduction of $\eta$ mesons}

\begin{figure}
\resizebox{0.48\textwidth}{!}{\includegraphics{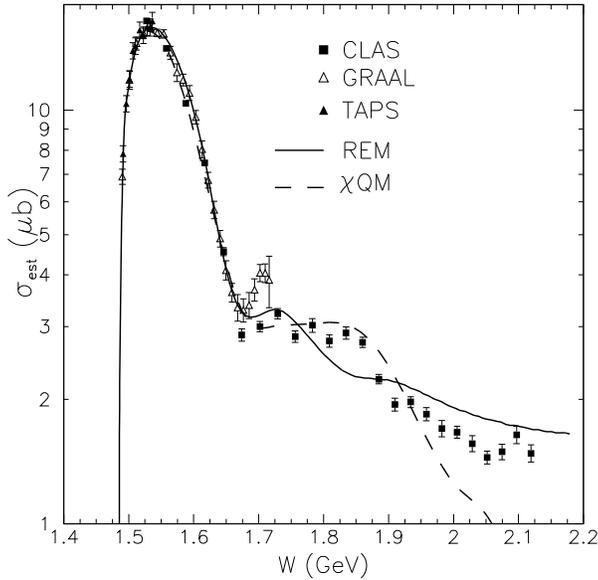}}
\caption{\small Total $\eta$ photoproduction cross sections from protons.}
\label{fig:clas_eta}
\end{figure}

New results on $\gamma p \rightarrow p \eta$  have recently become available 
from CLAS\cite{eta_prl} 
covering the resonance region for $W < 2.15$ GeV. Nearly complete 
angular distributions have been measured and the total cross section has been 
extracted. The total cross section data are shown in Figure \ref{fig:clas_eta}. The data show 
structure beyond the well known $N^*(1535)$ indicative of
 higher mass resonance contributions to the $p \eta$ channel. Further
analysis of the angular and energy dependences are needed to come to more
definite conclusions on the excitation of specific resonances.

\subsection{Resonances in virtual Compton scattering}

\begin{figure}
\resizebox{0.48\textwidth}{!}
{\includegraphics{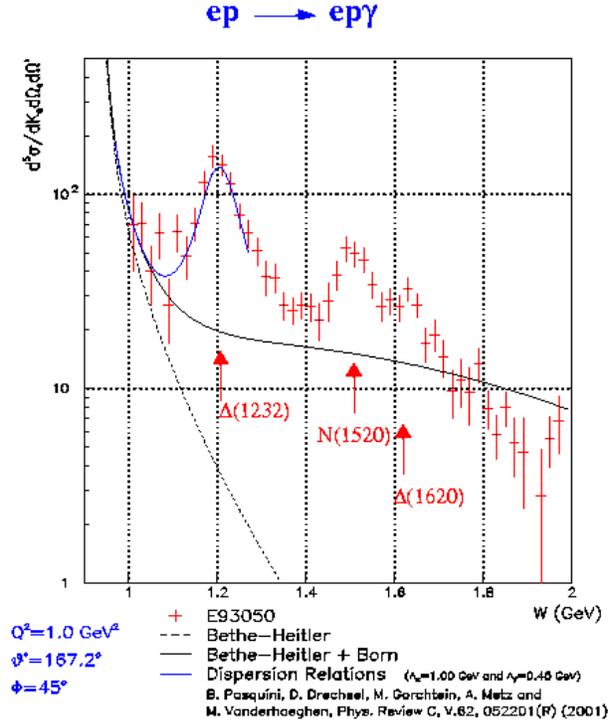}}
\caption{\small Differential cross section for virtual Compton scattering at 
$Q^2=1$ GeV$^2$. The final state photon is in the backward direction relative
to the virtual photon.}
\label{fig:vcs}
\end{figure}

Vitual Compton scattering, i.e. the process $\gamma^* p \rightarrow p \gamma $
is yet another tool in the study of excited baryon states. This process has 
recently been measured in experiment E93-050 in JLab Hall A\cite{fonvieille}
at backward photon angles. The excitation spectrum shown in Fig.\ref{fig:vcs}
exhibits clear resonance 
excitations at masses of known states such as the $\Delta(1232)$, $N^*(1520)$,
and $N^*(1650)$. The attractive feature of this process is the absence of
final state interaction which complicates the analysis of processes with 
mesons in the final state. The disadvantage is the low rate
 which makes it difficult to collect sufficient statistics for a 
full partial wave analysis.

\section{Baryon spectroscopy at short distances}

Inelastic virtual Compton scattering in the deep inelastic regime (DVSC) 
can provide a new avenue of resonance studies at the elementary quark level.
The process of interest is $\gamma^* p \rightarrow \gamma N^* (\Delta^*)$ 
where the virtual photon has high virtuality ($Q^2$).
The virtual photon couples to an elementary quark with longitudinal momentum 
fraction $x$, which is re-absorbed into the baryonic system with
momentum fraction $x-\xi$, after having emitted a high energy photon. 
The recoil baryon system may be a ground state proton or an excited state. 
The elastic DVCS process has recently been measured at JLab \cite{stepanya} 
and at DESY \cite{hermes} in polarized electron proton scattering, 
and the results are consistent with predictions from perturbative QCD and the
twist expansion for the process computed at the quark-gluon level. 
The theory is under control for small momentum transfer to the final state 
baryon. For the inelastic process, where a $N^*$ or 
$\Delta$ resonance is excited, the process can be used to study 
resonance transitions at the elementary quark level. Varying the parameter 
$\xi$  and the momentum transfer to the recoil baryon probes the two-parton 
correlation functions, or generalized parton distributions (GPDs). 

\begin{figure}
\resizebox{0.48\textwidth}{!}{\includegraphics{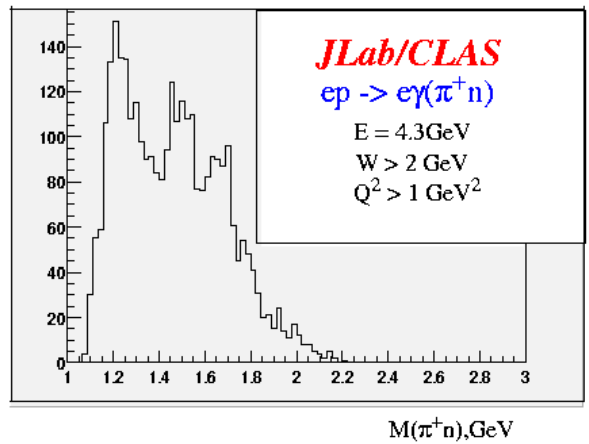}}
\caption{\small Inelastic deeply virtual Compton scattering measured 
in CLAS. The recoiling $(n\pi^+)$ system clearly shows the excitation of
several resonances, the $\Delta(1232)$, $N^*(1520)$, and $N^*(1680)$. }
\label{fig:deltadvcs}
\end{figure}

That this process is indeed present at a measureable level is seen in the 
preliminary data from CLAS \cite{guidal} shown in Fig. \ref{fig:deltadvcs}. 
The reaction is measured at invariant masses $W > 2$ GeV. 
The recoiling baryonic system clearly  
shows the excitation of resonances, the $\Delta(1232)$, $N^*(1520)$, 
and $N^*(1680)$. While these are well known states that are also excited in 
the usual s-channel processes, the DVCS process has the advantage 
that it decouples the photon virtuality $Q^2$ from the 4-momentum 
transfer to the baryon system. $Q^2$ may be chosen sufficiently high 
such that the virtual photon couples to an elementary quark, while the 
momentum transfer to the nucleon system can be varied independently from 
small to large values. In this way, a theoretical framework employing 
perturbative methods can be used to probe the ``soft'' $NN^*$ transition,
allowing to map out internal parton correlations for this transition.

\section{Conclusions} 
\label{sec:conclusions}
Electroexcitation of nucleon resonances has evolved to an 
effective tool in studying nucleon structure in the regime of 
strong QCD and confinement. The new data from JLab in the 
$\Delta_{3/2^+}(1232)$ and $N^*_{1/2^-}(1535)$ regions give a
consistent picture of the $Q^2$ evolution of the transition 
form factors. The $R_{EM}$ and $R_{SM}$ ratios for the  
$\gamma^* N \Delta(1232)$ transition are consistent with an 
oblate deformation of the $\Delta^+$. This is now
also confirmed by calculations in full lattice QCD. 
Large data sets in different channels including
polarization observables will vastly improve the analysis of
states such as the ``Roper'' $N^{\prime}_{1/2^+}(1440)$, and 
many other higher mass states. A preliminary analysis of $n\pi^+$ and 
$p \pi^{\circ}$ cross section data and beam polarization asymmetries 
at $Q^2 = 0.4$ GeV$^2$ show little indication of the 
 $N^{\prime}_{1/2^+}(1440)$, which is consistent with earlier analyses 
showing a fast drop of the Roper excitation strength  with $Q^2$.   
A strong resonance signal near 1.72 GeV, seen with CLAS in the 
$p\pi^+\pi^-$ channel, exhibits hadronic 
properties which appear incompatible with any of the known states 
in this mass region and may indicate a new $N^*_{3/2^+}(1720)$ state. 

While s-channel resonance excitation will 
remain the backbone of the $N^*$ program for years to come, 
inelastic deeply virtual Compton scattering is a promising new tool in 
resonance physics at the elementary parton level that allows the study 
of parton-parton correlations in resonance transition within a well defined 
theoretical framework.

\vspace{0.3cm}\noindent
The Southeastern Universities Research Association (SURA) operates the 
the Thomas Jefferson National Accelerator Facility for the United States
Department of Energy under Contract No. DE-AC05-84ER40150.

\end{document}